\documentclass[reprint,groupedaddress,amsmath,amssymb,aps,prl,floatfix]{revtex4-1}

\usepackage{amsmath,amssymb,graphicx,mathrsfs,amsfonts}

\draft 

\begin{document}

\title{Magnetically-Induced Optical Transparency on a Forbidden Transition in Strontium for Cavity-Enhanced Spectroscopy} 



\author{Matthew N. Winchester}
\affiliation{JILA, NIST, and Dept. of Physics, University of Colorado, 440 UCB, Boulder, CO  80309, USA}
\author{Matthew A. Norcia}
\affiliation{JILA, NIST, and Dept. of Physics, University of Colorado, 440 UCB, Boulder, CO  80309, USA}
\author{Julia R.K. Cline}
\affiliation{JILA, NIST, and Dept. of Physics, University of Colorado, 440 UCB, Boulder, CO  80309, USA}
\author{James K. Thompson}
\affiliation{JILA, NIST, and Dept. of Physics, University of Colorado, 440 UCB, Boulder, CO  80309, USA}

\date{\today}

\begin{abstract}
In this work we realize a narrow spectroscopic feature using a technique that we refer to as magnetically-induced optical transparency.  A cold ensemble of $^{88}$Sr atoms interacts with a single mode of a high-finesse optical cavity via the 7.5~kHz linewidth, spin forbidden $^1$S$_0$ to $^3$P$_1$ transition. By applying a magnetic field that shifts two excited state Zeeman levels, we open a transmission window through the cavity where the collective vacuum Rabi splitting due to a single level would create destructive interference for probe transmission. 
The spectroscopic feature approaches the atomic transition linewidth, which is much narrower than the cavity linewidth, and is highly immune to the reference cavity length fluctuations that limit current state-of-the-art laser frequency stability.


 
\end{abstract}


\maketitle 

There has been a dedicated effort in recent years to improve the frequency stability of lasers \cite{kessler2012sub} used to probe optical atomic clocks \cite{bloom2014optical,hinkley2013atomic,ushijima2015cryogenic}. Improvements in these precision measurement technologies are essential for advancing a broad range of scientific pursuits such as searching for variations in fundamental constants \cite{rosenband2008frequency}  , gravitational wave detection \cite{kolkowitz2016gravitational, graham2013new}, and physics beyond the standard model \cite{derevianko2014hunting,blatt2008new}. Associated improvements in atomic clocks would also  advance recent work on relativistic geodesy \cite{chou2010optical}. 

The frequency stability of current state-of-the-art lasers is limited by thermal fluctuations in the reference cavity mirror coatings, substrates, and spacer \cite{numata2004thermal}. This problem can be alleviated by creating systems that rely on an ensemble of atoms, rather than the reference cavity, to achieve stable optical coherence. Recent approaches include cavity-assisted non-linear spectroscopy \cite{MMT11,PhysRevLett.114.093002,PhysRevA.92.053820} and superradiant lasers \cite{bohnet2012steady,PhysRevX.6.011025,Norciae1601231, MYC09}. Both approaches use narrow forbidden transitions with linewidths ranging from 7.5 kHz to 1 mHz. These novel systems are absolute frequency references and are intrinsically less sensitive to both fundamental thermal and technical vibrations that create noise on the optical cavity's resonance frequency. 



Here we demonstrate a new linear spectroscopy approach in which a static magnetic field can induce optical transparency in the transmission spectrum of an optical cavity.   The center frequency of the transparency window is shown to be insensitive to changes in the cavity-resonance frequency and to first-order Zeeman shifts. 
The observed linewidth of the feature approaches the natural linewidth of the 7.5~kHz optical transition and can be insensitive to inhomogeneous broadening of the atomic transition frequency. The  linewidth of the feature is an important attribute for laser stabilization, as a laser stabilized to a narrow spectroscopic feature is less sensitive to technical offsets than a laser stabilized to a broader feature.   In the future, it might be possible to extend this technique to even narrower optical transitions for enhanced spectroscopic sensitivity in atoms such as calcium and magnesium.  

%

In partial analogy to electromagnetically induced transparency (EIT) \cite{boller1991observation,mucke2010electromagnetically,hernandez2007vacuum}, we refer to this effect as magnetically induced transparency (MIT).  In EIT, a control laser is used to create a variable-width transparency window for slowing light \cite{hau1999light}, for stopping light \cite{liu2001observation}, for quantum memories \cite{zhao2009long}, and even for creating effective photon-photon interactions \cite{peyronel2012quantum,dudin2012strongly,ningyuan2016observation}. It might be possible to utilize controlled magnetic fields and long-lived optical states to realize similar goals.

\begin{figure}[!htb]
\includegraphics[width=3.375in, ]{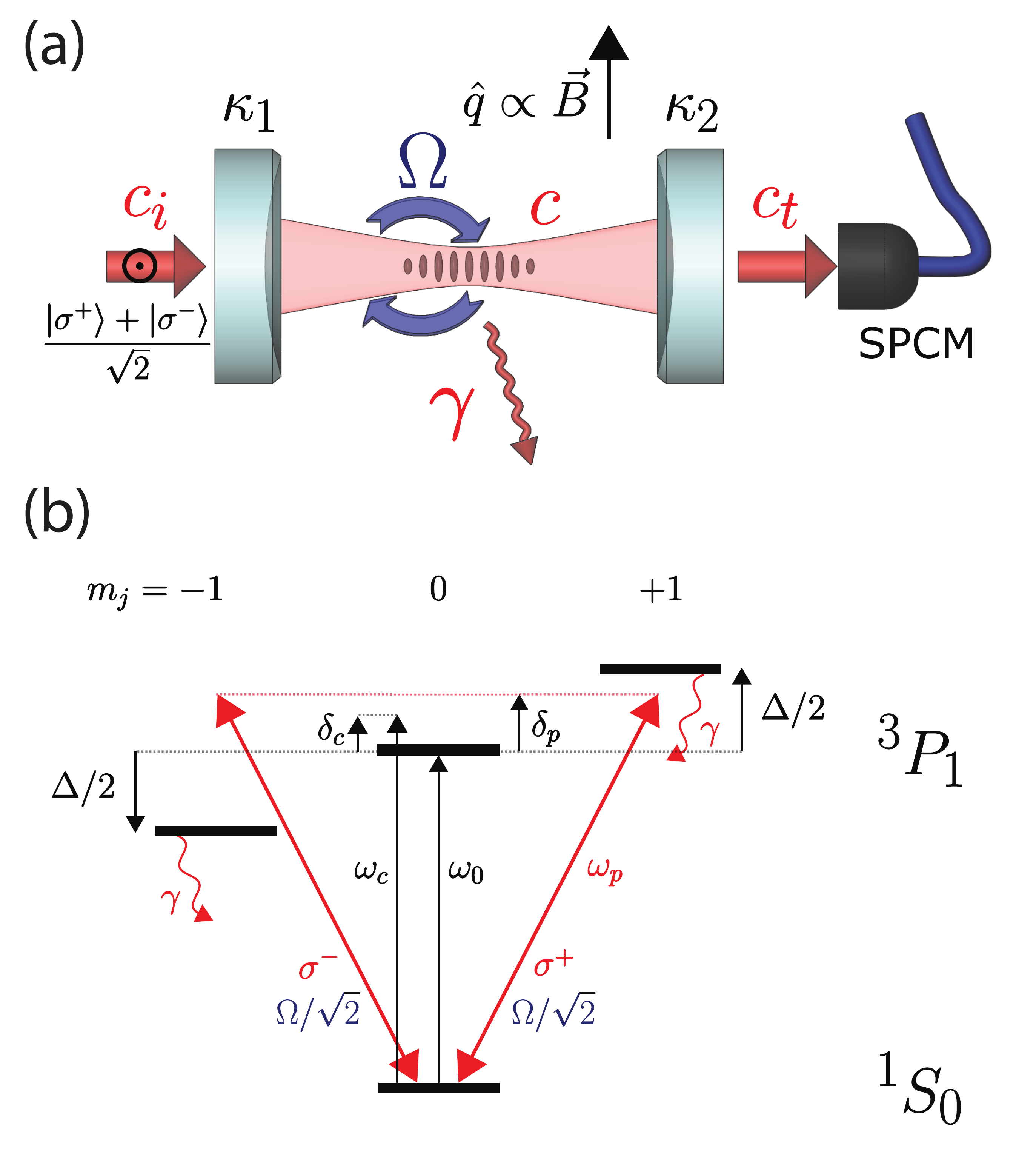}
\caption{(a) Simplified experimental diagram.  The quantization axis $\hat{q}$ is along the vertical magnetic field $\vec{B}$.  Horizontally polarized probe light with amplitude $c_i$ is a linear combination of left and right circularly polarized light. The light is transmitted through the input and output mirrors with characteristic rates $\kappa_1$ and $\kappa_2$ giving a cavity linewidth $\kappa=\kappa_1+\kappa_2$.  The light can be coherently absorbed by the atoms (brown ovals) and reemitted into the cavity (with field amplitude $c$) at the vacuum Rabi frequency $\Omega$. The atomic transition decays into free space at rate $\gamma/2\pi=7.5$~kHz. The transmitted field $c_t$ is detected on a single photon counting module (SPCM). (b) Corresponding energy level diagram. The applied magnetic field creates a Zeeman splitting $\Delta$ between the two states $^3P_1$, $m_j=\pm1$ that interact with the horizontally polarized light inside the cavity. Also shown are the the average atomic transition frequency $\omega_0$, the empty cavity resonance frequency $\omega_c$, and the detunings of the probe and cavity frequencies $\delta_p$ and $\delta_c$ relative to the atomic frequency $\omega_0$.  The $m_j=0$ state is shown, but does not interact with the horizontally polarized cavity-field.}
\label{fig:leveldiagram}
\end{figure}

In our experiment we create a strongly coupled atom-cavity system by loading up to $N=1.3\times10^{6}$ $^{88}$Sr atoms into a 1D optical lattice supported by a high-finesse optical cavity.  The peak trap depth is $100(10)~\mu$K and the atoms are laser-cooled to $10(1)~\mu$K (see Fig.~\ref{fig:leveldiagram}a and Refs.~\cite{norcia2015strong,PhysRevX.6.011025,Norciae1601231}). We tune a TEM00 resonance of the cavity at frequency $\omega_c$ to be near resonance with the dipole-forbidden singlet to triplet optical transition $^1$S$_0$ to $^3$P$_1$ at frequency $\omega_0$ or wavelength 689~nm (see Fig.~\ref{fig:leveldiagram}b). The excited state $^3$P$_1$ spontaneously decays back to the ground state at a rate $\gamma=2\pi\times7.5~$kHz, and the cavity power decays at rate $\kappa=2\pi\times150.3(4)~$kHz. In $^{88}$Sr, the absence of nuclear spin means that the $^1$S$_0$ ground state is unique, while the $^3$P$_1$ excited state has three Zeeman sublevels.  

\begin{figure}[!htb]
\includegraphics[width=3.375in, ]{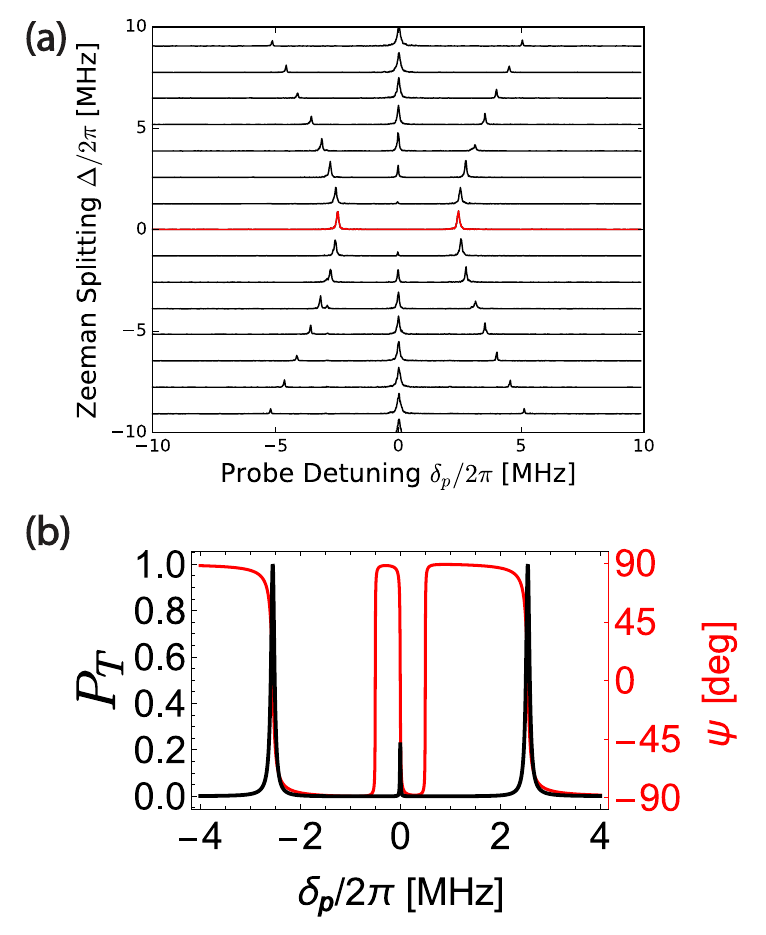}
\caption{(a) The transmitted power through the cavity versus the probe detuning $\delta_p$, with $\delta_c=0$. Each trace was taken for different applied magnetic fields, creating different Zeeman splittings $\Delta$ labeled on the vertical for each trace. The central red trace is taken for $\Delta=0$ and displays a collective vacuum Rabi splitting $\Omega/2\pi=5$~MHz.  When a magnetic field is applied perpendicular to the probe polarization, inducing a Zeeman splitting $\Delta$, a new transmission feature appears in between the two original resonances of the vacuum Rabi splitting.  
(b) Linearized theory showing the power $P_{T}$ and phase $\psi$ of the transmitted light, plotted here for $\Omega/2\pi=5$~MHz and $\Delta/2\pi=1$~MHz.}
\label{fig:waterfall}
\end{figure}

In the limit of zero applied magnetic field, our system responds to an applied probe as though each atom were a simple two-level system.  The ensemble can collectively absorb light from and then collectively reemit light into the cavity mode at the so-called collective vacuum Rabi frequency $\Omega= \sqrt{N} 2 g$. Here $2 g= 15$~kHz is the rms value of the single-atom vacuum Rabi frequency, which accounts for averaging over the standing-wave cavity mode.  This exchange creates two new transmission modes that are shifted away from the empty cavity's transmission peak by $\pm \Omega/2$, as shown by the central red trace in Fig.~\ref{fig:waterfall}a.

Because two orthogonal components of the probe light can couple to two distinct Zeeman sublevels, the coupled atom-cavity system has three normal modes of excitation, not two.  
In addition to the two modes at $\pm \Omega/2$ that lead to the transmission peaks, which we will refer to as the `bright' modes, there is a third `dark' mode whose frequency is equal to that of the atomic transition.  The dark mode is composed of an equal superposition of the two atomic excited states and a photonic component that vanishes as the magnetic field approaches zero.    




By applying a magnetic field $B$, we can mix photonic character into the dark mode, inducing transmission (also refered to as transparency) for probe light near $\omega_0$.  The probe light is horizontally polarized and is perpendicular to the vertically oriented magnetic field.  Each trace in Fig.~\ref{fig:waterfall}a corresponds to a different applied magnetic field with strength parameterized by the induced Zeeman frequency splitting $\Delta/2 \pi = (2.1$~MHz/G$) B$.

To describe the system, we extend the linearized input-output equations of \cite{PhysRevA.89.043837} to include an additional atomic transition written in a rotating frame at the average atomic transition frequency $\omega_0$ as:

\begin{eqnarray}
    \dot{a} &=&-\frac{1}{2}\left(\gamma + \imath \Delta \right) a -\imath \frac{1}{2 \sqrt{2}} \Omega \; c \\
    \dot{b} &=&-\frac{1}{2}\left(\gamma - \imath \Delta \right) b-\imath \frac{1}{2 \sqrt{2}} \Omega \; c  \\
    \dot{c} &=& -\frac{1}{2}\left(\kappa + \imath 2 \delta_c \right) c-\imath \frac{1}{2 \sqrt{2}} \Omega \left(a + b\right) + \sqrt{\kappa_1} c_i e^{\imath \delta_p t} .
\end{eqnarray}

\noindent Here, $\delta_c= \omega_c-\omega_0$ is the detuning of the cavity resonance frequency $\omega_c$ from atomic resonance, $\Omega$ is the observed collective vacuum Rabi splitting when $\Delta=0$, $\gamma$ is the decay rate of the excited atomic states, $\kappa$ is the cavity power decay rate, and $\kappa_1$ is the coupling of the input cavity mirror that is driven by an externally incident probe field with complex amplitude $c_i$ and at a probe frequency $\omega_p$ and detuning from atomic resonance $\delta_p=\omega_p-\omega_0$. The complex variables $a=\langle\hat{a}\rangle$, $b=\langle\hat{b}\rangle$, $c=\langle\hat{c}\rangle$ are expectation values of bosonic lowering operators describing the cavity $c$, and collective excitations of the two atomic transitions $a$ and $b$. The required Holstein-Primakoff approximation assumes weak excitation such that the number of atoms in the excited states $M_a=|a|^2$, $M_b=|b|^2\ll N $ is a small fraction of the total atom number $N$. The average number of photons in the cavity is given by $M_c=|c|^2$, and the complex field transmitted through the cavity is $c_t= \sqrt{\kappa_2} c$ with $\kappa_2$ the coupling of the output mirror. The transmitted probe power relative to incident probe power is $P_T= |c_t/c_i|^2$ and the relative phase is $\psi= \arg{(c_t/c_i)}$.


Figure~\ref{fig:waterfall}b shows the calculated steady-state transmitted power and phase for a single Zeeman splitting. The phase response changes rapidly near zero probe detuning, which results in a narrow MIT resonance compared to the broad vacuum Rabi splitting or bright modes for which the phase changes more slowly. 



In order to describe the linewidth of the dark state resonance, we introduce a mixing angle $\theta$ defined by $\sin^2\theta = \bar{\Delta}^2/(\Omega^2 + \bar{\Delta}^2)$. Here, the effective detuning is $\bar{\Delta}^2 = \Delta^2+\gamma^2$. The character of the dark state excitation is given by the ratio of the probability that the excitation is photonic-like $P_c= M_c/(M_c+M_a + M_b)= \sin^2\theta$ versus atomic-like  $P_{ab}= (M_a+M_b)/(M_c+M_a + M_b)= \cos^2\theta$. The dark state excitation can decay into free space at rate $R_{ab}$ or by emission through the cavity mirrors $R_c$, with the ratio of the rates given simply by $R_{ab}/R_c=\gamma/(\kappa \tan^2 \theta) = N C (\gamma/\bar{\Delta})^2$, where the single particle cooperativity parameter is $C= 4 g^2/\kappa\gamma$.


The linewidth of the dark state resonance can be written as: 

\begin{equation}
\kappa' = (\gamma \cos^2\theta  + \kappa \sin^2\theta)/b .
\end{equation}

\noindent The term in parentheses is a weighted average of atom and cavity linewidths that reflects the character of the mode.  The correction factor is $b = d \cos^2 \theta + \sin^2 \theta$, where $d= (\Delta^2-\gamma^2)/\bar{\Delta}^2$. When $\Delta \gg \gamma$, both $b$ and $d$ approach unity.  At small detunings $\Delta\sim\gamma$, the responses of the dark and bright modes to the applied drive become comparable, causing a modification of the correction factor. In the regime experimentally explored here ($b \approx 1$), $\kappa'$ is simply the full width at half maximum linewidth of the power transmission feature.  To define a linewidth valid in general, we define the linewidth via $\kappa'= 2 \left(\mathrm{d} \psi/\mathrm{d}\delta_p\right)^{-1}|_{\delta_c=\delta_p=0}$. For $\Omega \gg \Delta \gg \gamma$, the mixing angle is small and the linewidth approaches the atomic linewidth $\kappa'\approx \gamma$, which can be much narrower than the cavity linewidth $\kappa$.

\begin{figure}[!htb]
\includegraphics[width=3.375in, ]{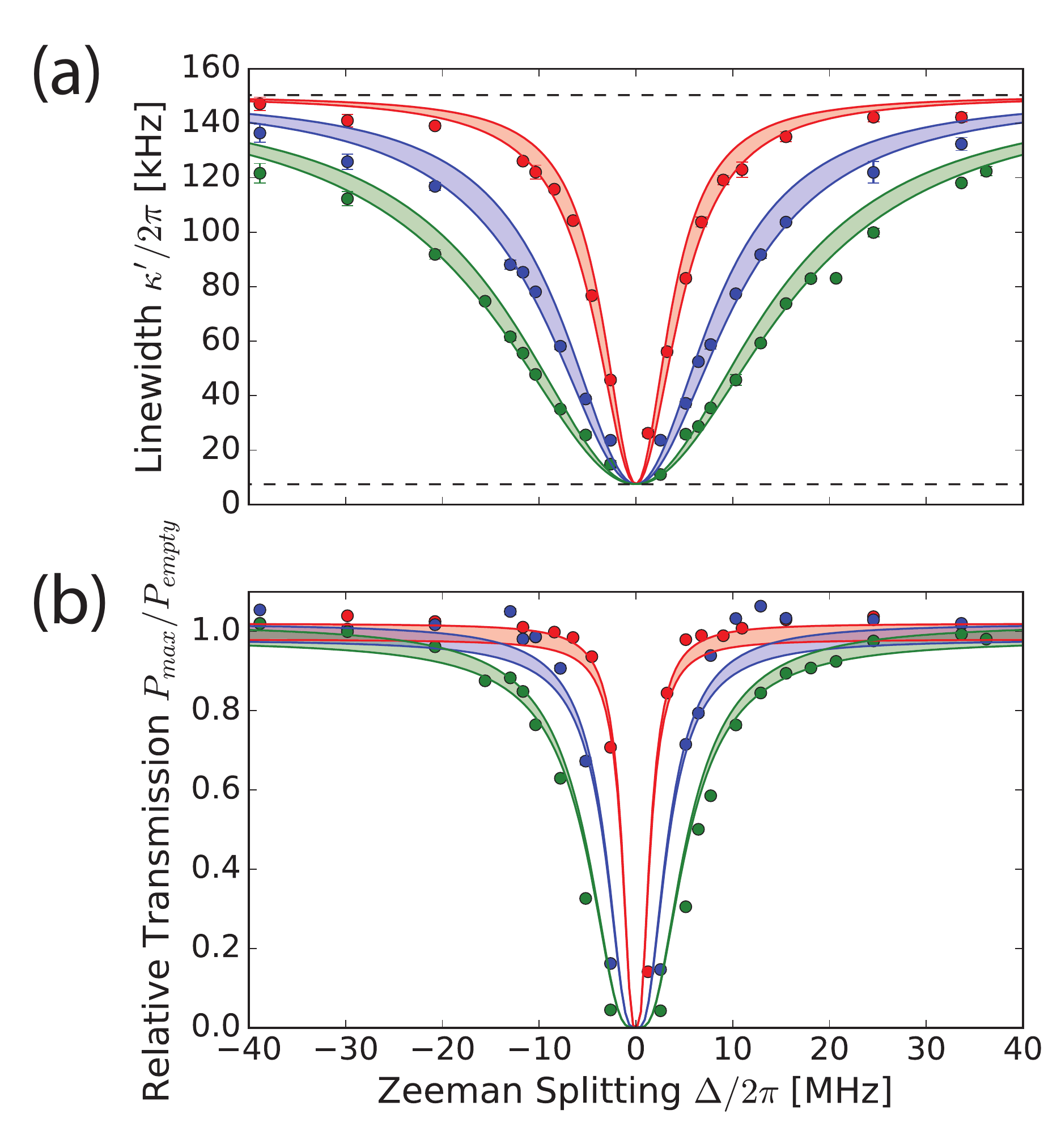}
\caption{(a) The measured linewidth of the central MIT transmission feature versus the induced Zeeman splitting between excited states.  The traces are taken for three different collective vacuum Rabi frequencies $\Omega/2\pi =$ 4.6(5) (red), 10(1) (blue), and 16(1) (green)~MHz, with values set by changing the total atom number $N$.  The upper dashed line is the empty cavity's linewidth $\kappa$, and the lower dashed line is the atomic transition's linewidth $\gamma$.  The minimum observed linewidth was 11~kHz.  The shaded regions are no-free parameter predictions from the linearized model introduced in the text, indicating the $\pm1$ standard deviation uncertainty bands based on independent measurements of $\Omega$. (b) The measured peak transmitted power of the central MIT transmission feature for the same collective Rabi frequencies. Here, the transmitted power is normalized to the peak transmitted power when the cavity is empty. Again the shaded regions indicate the $\pm1$ standard deviation uncertainty bands for the predictions.}
\label{fig:linewidth}
\end{figure}

We measure the linewidth for the central dark resonance by linearly sweeping the probe laser's frequency over the cavity resonance and recording a time-trace of power transmitted on a single photon counting module. A Lorentzian is fit to the central feature to extract the full width at half maximum. This measurement is taken for a range of different Zeeman splittings and vacuum Rabi splittings by varying the applied dc magnetic field and atom number respectively. Figure \ref{fig:linewidth}a shows collected data plotted against the theoretical prediction at several Rabi frequencies. For very small $\Delta$ the feature becomes increasingly narrow, approaching the atomic transition linewidth $\gamma=2\pi\times7.5~$kHz. For very large $\Delta$ the feature linewidth approaches the cavity decay rate $\kappa$.

Figure \ref{fig:linewidth}b shows the peak transmitted power at the MIT feature's resonance for the same data show in Fig.~\ref{fig:linewidth}a. The linearized theory predicts that the peak transmitted power is given by:


\begin{equation}
P_{max}=\frac{4 \kappa_1 \kappa_2}{\kappa^2} \frac{1}{\left(1+\frac{\gamma}{\kappa \tan^2\theta}\right)^2} .
\end{equation}

\noindent Note that the term in the denominator above is just the ratio of excitation decay rates $R_{ab}/R_c$. For large detunings $\Delta\gg \Omega, \gamma$, the peak transmission goes to that of an empty cavity $P_{max}\rightarrow P_{empty}=4 \kappa_1\kappa_2/\kappa^2$.   




In the regime experimentally explored here, a change in the cavity resonance frequency $\omega_c$ by $\Delta \omega_c$ leads to a change in the dark state resonance frequency $\omega_{D}$ by a much smaller amount $\Delta\omega_{D}$. The pulling coefficient $P=\Delta \omega_{D}/\Delta \omega_{c}$ expresses this ratio. A small pulling coefficient $P\ll1$ is desirable for a frequency reference as it will be less sensitive to thermal fluctuations and technical noise on the reference cavity. We can extract a pulling coefficient applicable to all parameter regimes from the linearized theory by considering how much the probe and cavity detunings would have to change to create equal changes in the quadrature amplitude of the transmitted field (such as one might measure using homodyne detection). This general pulling coefficient can be expressed as:    




\begin{equation}
    P=\frac{\sin^2 \theta}{b} \, .
\end{equation}

\noindent In the typical regime of operation ($b \approx 1$), this is simply the cavity-like fraction of the dark excitation.  

\begin{figure}[!htb]
\includegraphics[width=3.375in, ]{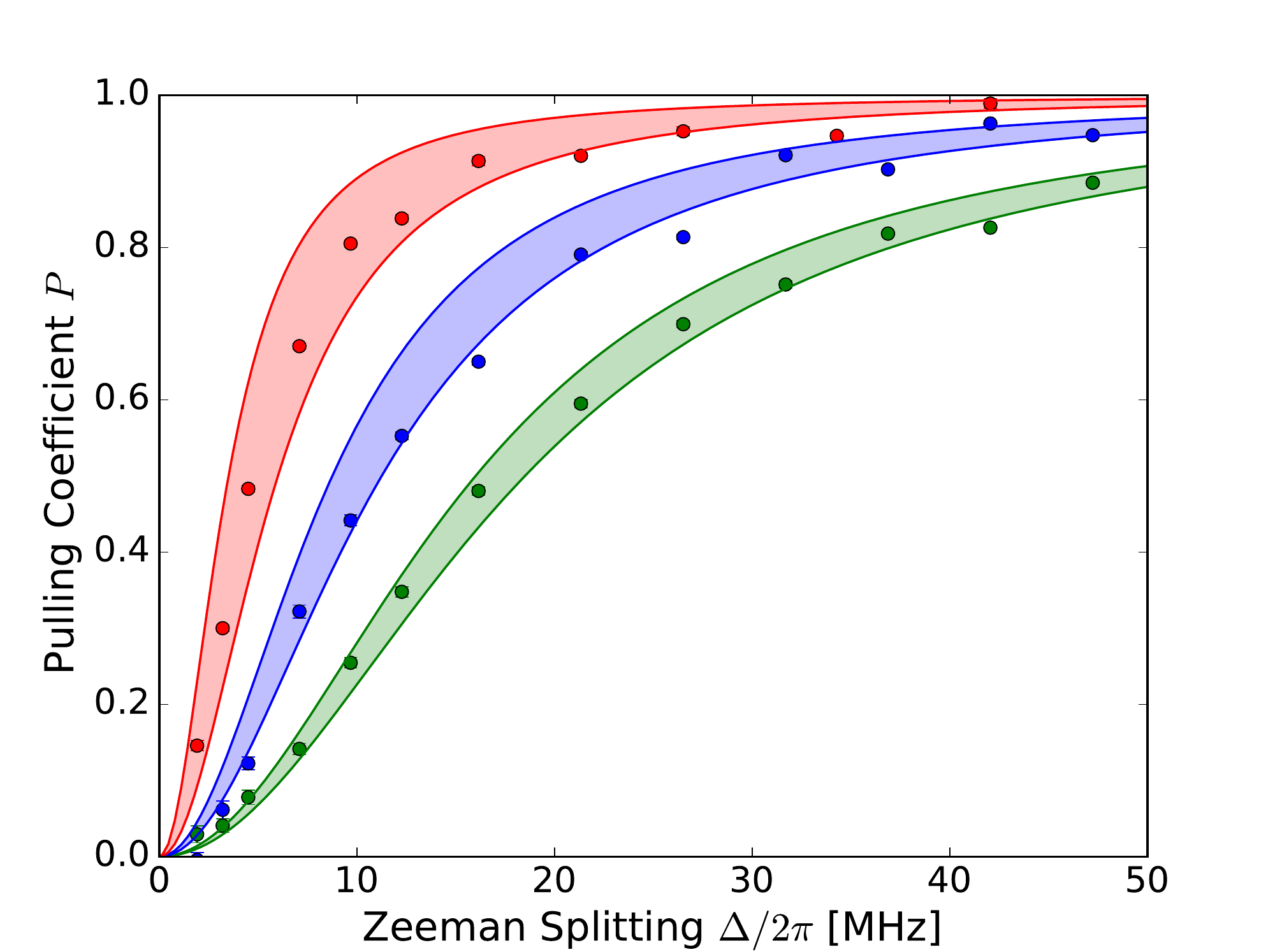}
\caption{The pulling coefficient $P$ versus Zeeman splitting $\Delta$ for several collective vacuum Rabi frequencies $\Omega/2\pi= 5(1)$ (red), 10(1) (blue), and 17(1) (green) MHz.  The prediction from the linearized theory is shown with $\pm1$ standard deviation bands.}
\label{fig:pulling}
\end{figure}

In Fig.~\ref{fig:pulling}, we show the measured pulling coefficient versus splitting $\Delta$ for several values of $\Omega$, along with the predicted pulling coefficients from the linearized theory. The pulling coefficients were measured by sweeping the probe laser frequency across the dark resonance and fitting the center frequency $\omega_{D}$ with a Lorentzian fit model.  
This is then repeated while toggling $\omega_c$ between two values separated by 100~kHz, and the pulling coefficient is determined from the change in $\omega_{D}$ versus $\omega_c$. Our lowest measured pulling coefficients are below $P=0.05$. 

In principle, $\Delta$ can be reduced further to reach a smaller pulling coefficient, at the expense of transmitted power.  
The theoretical pulling coefficient reaches a minimum value of $P = 1/(1+ N / (8 M_c ))$ at $\Delta = \sqrt{3}\gamma$, where $M_c= (\gamma/(2 g))^2$ is the so-called critical photon number.  The critical photon number is proportional to the cavity mode volume and atomic linewidth, but does not depend on the mirror reflectivity.   As a result, small pulling coefficients are reached by working with small cavity volumes and very narrow linewidth transitions.  For spectroscopic applications, one would want to balance the desire for a low pulling coefficient against the need to collect transmitted photons without inducing heating in the atomic ensemble due to free-space scattering.  The optimal parameter regime will depend on the specific requirements of the system.

While the majority of this work has been done with the atoms trapped in the Lamb-Dicke regime (i.e.~confined to much less than the wavelength of the probe light) with respect to the cavity axis,  we have also performed scans of the cavity transmission spectrum in which the atoms were unconfined along the cavity axis.  In this configuration, the rms Doppler shift along the cavity axis is roughly 45~kHz.  Despite this inhomogeneous broadening, we observe a center feature linewidth of 18.5~kHz, which we believe is limited by technical noise on the cavity frequency that arises when we turn down the lattice depth to release the atoms.  We expect the linewidth of the dark feature to be insensitive to inhomogeneous broadening so long as $\Delta$ is much larger than the inhomogeneous broadening \cite{PhysRevA.53.2711}. This insensitivity to Doppler broadening may make such techniques suitable to continuously operating atomic beam experiments, where confining the atoms to the Lamb-Dicke regime would be challenging.  

To summarize, we have demonstrated a technique to realize a narrow spectroscopic feature based on collective interaction between an ensemble of atoms and a high finesse optical cavity.  The center frequency of the feature can be made highly insensitive to changes in cavity resonance frequency.  In analogy to EIT, this technique may also be applicable to tasks relevant for information processing, especially if applied to an optical transition with even narrower linewidth.

We acknowledge contributions to the experimental apparatus
by Karl Mayer. All authors acknowledge financial support from DARPA QuASAR, ARO, NSF PFC, and NIST. J.R.K.C. acknowledges financial support from NSF GRFP. This work is supported by the National Science Foundation under Grant Number 1125844.

\bibliography{ThompsonLab.bib}

\end{document}